\documentclass[prd,twocolumn,nofootinbib,preprintnumbers,floatfix]{revtex4}  
\usepackage[plainpages=false,  colorlinks=true, anchorcolor=blue, linkcolor=blue, citecolor=blue, bookmarks=false]{hyperref}
\usepackage{graphicx}  
\usepackage{dcolumn}   
\usepackage{bm}        

\usepackage{natbib}
\usepackage{amsmath}   
\usepackage{longtable}  
\usepackage{float}    
\usepackage{tasks}

\usepackage{amssymb} 

\newcommand{\rthis}[1]{\textcolor{black}{#1}}

\begin{document}
\newcommand{\apjl}{Astrophys. J. Lett.}
\newcommand{\apjs}{Astrophys. J. Suppl. Ser.}
\newcommand{\aap}{Astron. \& Astrophys.}
\newcommand{\aj}{Astron. J.}
\newcommand{\pasp}{PASP}
\newcommand{\araa}{Ann. Rev. Astron. Astrophys. } 
\newcommand{\aapr}{Astronomy and Astrophysics Review}
\newcommand{\ssr}{Space Science Reviews}
\newcommand{\mnras}{Mon. Not. R. Astron. Soc.}
\newcommand{\apss} {Astrophys. and Space Science}
\newcommand{\jcap}{JCAP}
\newcommand{\pasj}{PASJ}
\newcommand{\pasa}{Pub. Astro. Soc. Aust.}
\newcommand{\physrep}{Physics Reports}



\title{A test of cosmic distance duality relation using SPT-SZ  galaxy clusters, Type Ia supernovae, and cosmic chronometers}

\author{Kamal \surname{Bora}}
 \altaffiliation{E-mail: ph18resch11003@iith.ac.in}

\author{Shantanu \surname{Desai}}%
\altaffiliation{E-mail: shntn05@gmail.com}

\date{\today}

\begin{abstract}
We carry out a test of the cosmic distance duality relation using a sample of 52 SPT-SZ clusters, along with  X-ray measurements from XMM-Newton. To carry out this test, we need an estimate of the luminosity distance ($D_L$) at the redshift of the cluster. For this purpose, we use three independent methods: directly using $D_L$ from  the closest  Type Ia Supernovae from the Union 2.1 sample,  non-parametric reconstruction of  $D_L$ using   the same Union 2.1 sample, and finally  using  $H(z)$ measurements from cosmic chronometers and reconstructing $D_L$ using Gaussian Process regression.  We use four different functions to characterize the deviations from CDDR. All our results for these ($4 \times  3$) analyses are consistent with CDDR to within 1$\sigma$.
 
\end{abstract}

\affiliation{Department of Physics, Indian Institute of Technology, Hyderabad, Kandi, Telangana-502285, India}

\maketitle

\section{Introduction}

One of the most fundamental tenet in Cosmology  is the cosmic distance duality relation (CDDR, hereafter)  between the angular diameter distance ($D_A$) and the luminosity distance ($D_L$),  which is sometimes also known as the  Etherington relation~\cite{etherington,Ellis07}. This relation is given by:
\begin{equation}
    \eta(z) \equiv \frac{D_L}{D_A (1+z)^2}=1
    \label{eq:cddr}
\end{equation}
The CDDR involves three assumptions~\cite{bassett04,Melia21}: spacetime is endowed with a metric theory of gravity; photons propagate along null geodesics; and the total number of photons is conserved. The CDDR relation is one of the edifice of the standard $\Lambda$CDM model~\cite{Ratra08}. Although, the $\Lambda$CDM  is very successful in describing the large scale structure from CMB and other probes~\citep{planck18}, a number of tensions have crept up in recent years, such as the Hubble constant tension between low redshift and high redshift probes~\cite{Divalentino21,Verde,Bethapudi}, $\sigma_8$ tension~\cite{Benisty}, Lithium-7 problem in Big-Bang Nucleosynthesis~\cite{Fields}, core-cusp and missing satellites problem~\cite{Bullock,Weinberg}, failure to detect cold dark matter in the laboratory~\cite{Merritt}, failure to explain regularities in cosmic structure at galactic scales such as the radial acceleration relation and Baryonic Tully-Fisher relation~\cite{Mcgaugh16,Pradyumna}.
Therefore, a large number of works have tried to test the CDDR in a model-independent way using a variety of astrophysical probes~\cite{uzan04,bernardis06,khedekar11,li11,nair11,meng12,ellis13,liao16,lv16,li18,lin18,YangSL,ruan18,lyu20,Melia21,arjona20}. Any violation of CDDR would result in a value for $\eta(z)$ in Eq.~\ref{eq:cddr}, different from one. In such cases, $D_L$ can be expressed in terms of $D_A$ and $\eta(z)$, which is sometimes known as the deformed CDDR relation~\cite{holanda19}. 

In this work, we shall use galaxy cluster observations observed in the microwave, via Sunyaev-Zeldovich effect (SZ, hereafter) and X-rays to test CDDR. Galaxy clusters are the most massive virialized collapsed objects in the universe and are wonderful laboratories for a whole range of topics in Cosmology, galaxy evolution,  and fundamental Physics~\cite{allen11,borgani12,Vikhlininrev,Desai}.
The first test of CDDR using galaxy clusters was done by~\cite{uzan04}. They used the X-ray and SZ observations from \cite{reese02} to test the validity of the CDDR,  and obtained  a value for $\eta$ consistent with  one at $2\sigma$. Later, \citet{bernardis06} did a similar test using the angular diameter distances to 38 galaxy clusters from X-ray observations with Chandra  and SZ observations with the OVRO and BIMA interferometric arrays. Their results were consistent with  no violation of CDDR at $1\sigma$.
The first model-independent cosmological test of CDDR was carried out in ~\cite{holanda10}, by using joint SZ  and X-ray surface brightness measurements, along with Type Ia supernovae (SNe)  from the Constitution data~\cite{hicken09}.  Here, they used two different sample of galaxy clusters from~\cite{laroque06} and ~\cite{ettori09}, corresponding to  elliptical and spherical geometry, respectively. 
For their analyses,  two different parametric forms for $\eta(z)$ were tested. Their results for clusters with elliptical geometry were consistent with CDDR to within $2\sigma$. For the sample with spherical geometry,  discrepancy with CDDR at about $3\sigma$ was reported.
Hence, they concluded that the elliptical geometry of galaxy clusters is compatible with the validity of CDDR but not the spherical geometry.
\citet{goncalves12} used X-ray $f_{gas}$ data, along with SNe Ia from the Union 2 compilation~\cite{Amanullah}. Here also, the same two samples of galaxy clusters~\cite{laroque06,ettori09} were used. This  work found that one of the samples~\cite{laroque06} is consistent with CDDR, whereas the other sample~\cite{ettori09} shows a 3.5$\sigma$ deviation from CDDR.
Later,~\citet{holanda12} reported no violation of CDDR  using  a combination of SZ and X-ray gas mass fraction data~\cite{laroque06}.
\citet{liang13} tested CDDR using  a sample of 38 angular diameter distances  from galaxy clusters assuming the spherical model~\cite{Bona}, along with SNe Ia  data from the Union 2 compilation~\citep{Amanullah}. 
They found no evidence for the violation of CDDR at $1-2\sigma$, depending on the parameterization used. 
\citet{yang13} carried out a model-independent test of CDDR using the angular diameter distances from two different cluster samples~\cite{Bona,Defillips} corresponding to spherical and elliptical geometry, respectively, in conjunction with the Union 2 SNe Ia sample~\cite{Amanullah}. However, instead of using the SNe Ia  $D_L$ from the distance modulus, they used the original SNe data for the  rest-frame  peak magnitudes and other parameters  describing the influence of intrinsic color and reddening by dust. They obtained no violation of CDDR at $1\sigma$ for
both the elliptical and spherical cluster samples, but concluded that the spherical model~\cite{Bona} can better describe the intrinsic geometry of clusters as compared to the elliptical one~\cite{Defillips}, if the CDDR relation is valid.
\citet{costa15} used two different methods to test CDDR using   38 angular diameter distances obtained from ~\cite{Bona} (which was also used in earlier tests of CDDR~\cite{yang13}),  along with  $f_{gas}$   measurements~\cite{laroque06}  spanning the redshift range $0.14 < z < 0.89$.  This was then compared with an independent estimate of the angular diameter  distance, reconstructed using $H(z)$ data from cosmic chronometers and BAO. In this work, the departure from CDDR was reconstructed in a non-parametric manner. Both the methods were consistent with  the validity of CDDR at $1\sigma$.~\citet{holanda16} studied the temporal variation of the  fine structure constant, modified distance duality relation, and the modified evolution law of the cosmic microwave background radiation by using the angular diameter distances to 29 galaxy clusters~\cite{Bona}, Union 2.1  Type Ia SNe compilation~\cite{suzuki12}, and 38 $T_{CMB}$ measurements~\cite{Luzzi,Hurier}.  All of  these variants/modifications of the Standard Model arise naturally  from a class of modified gravity  theories, which break the Einstein Equivalence Principle. Their results are consistent with no violation of these laws at $1\sigma$. (See  also ~\cite{holanda1611,holanda1705} for similar follow-up studies).
~\citet{holanda19} (H19, hereafter) have also done a study of the validity of  CDDR relation  using the $Y_{SZ}-Y_X$ scaling relation for 61 Planck ESZ clusters jointly with deep XMM-Newton archive observations in the redshift range $0.044 \leqslant z \leqslant 0.444$~\cite{PlanckESZ}, along with Type Ia SNe from the Pantheon compilation~\cite{Pantheon}. They reported  no violation of CDDR at $1.5\sigma$. Most recently, \citet{silva20} did a Bayesian model comparison assuming CDDR violation for different forms of $\eta(z)$ using two sets of data namely SNe Ia~\cite{Pantheon} + angular diameter distances from galaxy clusters~\cite{Defillips} and SNe Ia~\cite{Pantheon} + gas mass fraction data~\cite{mantz14}. Their results showed agreement with CDDR at $2\sigma$ for SNe Ia and the angular diameter distances from galaxy clusters sample, and at $1\sigma$ for SNe Ia along with the gas mass fraction measurements.

In this work, we shall use SZ selected clusters from the South Pole telescope in conjunction with X-ray followup observations from the  XMM-Newton telescope to test for a violation of CDDR. 
This paper is structured as follows. In Sec.~\ref{sec:methodology} , we briefly explain the basic theory behind the SZ and X-ray  observables used to test CDDR. In Sec.~\ref{sec:sample} , we describe our cluster sample. Sec.~\ref{sec:calc of $D_L$} discusses  the calculation of $D_L$ using three independent methods. Details  of our analysis and results can be found in Sec.~\ref{sec: analysis and results}. We conclude in Sec.~\ref{sec:conclusion}.

\section{methodology}
\label{sec:methodology}

The Inverse Compton scattering between the CMB photons and the hot electrons present in the intra-cluster medium causes a spectral distortion of the CMB black body spectrum~\cite{sz72,birki99,carl02,mrocz19}. This spectral distortion is known as the  SZ effect. It is nearly independent of redshift, and  hence can detect  galaxy clusters upto very high redshifts.
The distortion is characterized by a parameter, called Compton-$y$ parameter, which is given by~\cite{carl02,birki99}:
\begin{equation}
    y = \frac{{\sigma_T}k_B}{m_e c^2}  \int n_e T dl, 
\label{eq:y}    
\end{equation}
where,  $T$ is the electron temperature, $m_e$ is mass of the electron, $c$ is the speed of light, $n_e$ is the electron number density, $k_B$ is the Boltzmann constant, and $\sigma_T$ is the Thompson scattering cross section which is given by,  
\begin{equation}
    \sigma_T = \frac{8\pi}{3} \left(\frac{\epsilon^2}{m_e c^2}\right)^2  = \frac{8\pi}{3}  \left(\frac{\hbar^2\alpha^2}{m_{e}^2 c^2}\right)
\label{eq:sigmat}    
\end{equation}

If we model the  variation in $\alpha$  as $\alpha(z) \equiv \alpha_{0} \phi(z)$, where $\alpha_{0}$ is the present value of $\alpha$, the fractional variation in $\alpha$ can be written as,
\begin{equation}
\frac{\Delta \alpha}{\alpha_0}= \phi(z)-1
\label{eq:deltaalpha}
\end{equation}

We assume that the intra-cluster medium is an  ideal gas, with its equation of state  given by $P=n_e k_B T$, where $P$ refers the pressure of the intra-cluster gas. The relation between the integrated Compton parameter over the solid angle ($Y_{SZ}$) and the angular diameter distance to the cluster ($D_A$) is given by~\cite{bora21}:
\begin{equation}
    Y_{SZ} D_A^2  \propto \phi(z)^2
    \label{eq:YszDa2}
\end{equation}

The hot intra-cluster gas emits X-rays mainly through thermal bremsstrahlung emission~\cite{Sarazin,allen11}. If $M_g(R)$ is the gas mass within radius $R$ and $T_X$ is the X-ray temperature, then the average thermal energy of the cluster gas is given by~\cite{Kravtsov06},

\begin{equation}
    Y_X = M_g(R) T_X
    \label{eq:yxmg}
\end{equation}
An exact expression for the gas mass $M_{g} (<R)$ within a radius $R$ can be obtained by assuming that the number density obeys the spherical-$\beta$ model~\cite{Sarazin} and X-ray emission via thermal bremsstrahlung, and   is given by ~\cite{Sasaki}
\begin{widetext}
\begin{equation}
M_{g} (<R) = m_H \left( \frac{3 \pi \hbar m_e c^2}{2 (1+X) e^6}
\right)^{1/2}  \left( \frac{3 m_e c^2}{2 \pi k_B T_e} \right)^{1/4}
 \frac{1}{[\overline{g_B}(T_e)]^{1/2}}
{r_c}^{3/2} \left
[ \frac{I_M (R/r_c, \beta)}{I_L^{1/2} (R/r_c, \beta)} \right] [L_X
(<R)]^{1/2}\;,
\end{equation}
\end{widetext}
We then absorb the terms involving $e$, $c$, and $\hbar$ into $\alpha$, and rewrite this expression in terms of $\alpha \equiv \alpha_0 \phi(z)$ (from Eq.~\ref{eq:deltaalpha}) as follows:
\begin{widetext}
\begin{equation}
M_{g} (<R) = \alpha_0\phi(z)^{-3/2} m_H \left( \frac{3 \pi  m_e }{2 (1+X) \hbar^2}
\right)^{1/2}  \left( \frac{3 m_e}{2 \pi k_B T_e} \right)^{1/4}
 \frac{1}{[\overline{g_B}(T_e)]^{1/2}}
{r_c}^{3/2} \left
[ \frac{I_M (R/r_c, \beta)}{I_L^{1/2} (R/r_c, \beta)} \right] [L_X
(<R)]^{1/2}\;,
\label{eq:mgas}
\end{equation}
\end{widetext}
In Eq.~\ref{eq:mgas}, $L_X(<R)$ is the total X-ray luminosity; $X$ is the mass fraction of hydrogen, $r_c$ indicates the core radius; $T_e$ is the gas temperature, $\overline{g_B}(T_e)$ the 
Gaunt factor; and $I_M$,
$I_L$ are defined as follows~\cite{Sasaki}:
$$I_M (R/r_c, \beta) \equiv \int_0^{R/r_c} (1+x^2)^{-3 \beta/2} x^2 dx\;,$$
$$I_L (R/r_c, \beta) \equiv \int_0^{R/r_c} (1+x^2)^{-3 \beta} x^2 dx\;.$$
$L_X$, $r_c$, and $R$ are not direct observables, but depend on the underlying cosmological model and are given as~\cite{Sasaki}:
\begin{eqnarray}
\label{eq:Lx}
L_X (<R) &=& 4 \pi D_L^2 f_X(<\theta), \\ 
r_c &=& \theta_c D_A , \label{eq:rc} \\
R &=& \theta D_A ,
\end{eqnarray}
where $f_X(<\theta)$ is the total bolometric flux within the outer angular radius $\theta$, and  $\theta_c$ is the core angular radius. 
Therefore, from Eq.~\ref{eq:mgas}, Eq.~\ref{eq:Lx}, and Eq.~\ref{eq:rc}
we find that   $M_g(R)$ scales with $\phi(z)$, $D_L$, and $D_A$ according to:
\begin{equation}
      M_g (<\theta) \propto \phi(z)^{-3/2} D_L D_A^{3/2},
      \label{eq:mg2}
\end{equation}
 $D_L$ and $D_A$ are connected via CDDR (cf. Eq.~\ref{eq:cddr}). If we parameterize  a violation of CDDR  using the deformed CDDR  $D_L \equiv \eta(z) (1+z)^2 D_A$, then $M_g$ and consequently, $Y_X$ scales with $\alpha$ and $\eta(z)$ as~\cite{colaco19,bora21},
\begin{equation}
        Y_X   \propto M_g \propto \phi(z)^{-3/2} \eta(z)
        \label{eq:Yx}
\end{equation}

For a  wide class of modified theories of gravity~\cite{hees14,mina14,goncalves20},  Einstein's equivalence principle breaks down due to the coupling of the scalar field with the electromagnetic sector. In these theories, $\alpha(z)$ (and thereby $\phi(z)$) and $\eta(z)$ are intertwined and  related according to $\phi(z) = \left(\eta(z)\right)^2$.
$Y_{SZ}$ and $Y_{X}$  can then be rewritten in terms of $\eta(z)$  as:
\begin{equation}
    Y_{SZ} D_A^2  \propto \eta(z)^4
    \label{eq:Y_SZDa2}
\end{equation}
and
\begin{equation}
        Y_X   \propto  \eta(z)^{-2}
        \label{eq:Y_x}
\end{equation}
$Y_{SZ}$ and $Y_X$ are two different proxies for the thermal energy of the cluster~\cite{More}, so the ratio $Y_{SZ} D_A^2/C_{XSZ}Y_X$ is expected be constant with redshift, since  both these quantities scale   with redshift and mass in exactly the same way.  Simulations show that this  ratio is constant with $5-15\%$ scatter~\cite{planelles17,biffi14,fabjan11,kay12,stanek10}. The ratio would be exactly one for clusters with isothermal or universal temperature profiles~\cite{loken02,galli,colaco19}. 
From Eq.~\ref{eq:Y_SZDa2} and ~\ref{eq:Y_x}, for any violations of CDDR,  this ratio scales with $\eta(z)$ as:  
\begin{equation}
    \frac{Y_{SZ} D_A^2}{Y_X C_{XSZ}} =  C \eta(z)^6
\end{equation}
where $C_{XSZ} \approx 1.416 \times 10^{-19} \left( \frac{ Mpc^2}{M_\odot keV}\right)$ and $C$ is an arbitrary constant, which contains all the cluster astrophysics and is equal to one for an isothermal profile~\cite{colaco19}.

From Eq.~\ref{eq:yxmg} and~\ref{eq:mg2}, one can see that $Y_X$ scales with $D_A$ according $Y_X \propto D_A^{5/2}$,  if we assume a flat $\Lambda$CDM cosmology and the validity of CDDR. Following H19, we multiply  $Y_X$ by $D_A^{5/2} / (D_A^{ref})^{5/2}$, where $D_A^{ref}$ is the angular diameter distance in the fiducial reference $\Lambda$CDM model, corresponding to $\Omega_M=0.3$ and $\Omega_{\Lambda}=0.7$.
This  eliminates the dependence of $M_g(R)$, with respect to the fiducial cosmology. 
Hence we get,
\begin{equation}
    \frac{Y_{SZ} D_A^2 (D_A^{ref})^{5/2}}{Y_X C_{XSZ} D_A^{5/2}} =  C \eta(z)^6
\label{eq:ratiowitheta}
\end{equation}
Since we want to carry out a  test of CDDR in a model-agnostic fashion, we recast this equation in terms of $D_L$ by parameterizing any violation of CDDR using $\eta(z)$ defined in Eq.~\ref{eq:cddr}.
So Eq.~\ref{eq:ratiowitheta} can be written as,
\begin{equation}
    \frac{Y_{SZ}(D_A^{ref})^{5/2} (1+z)}{Y_X C_{XSZ} D_L^{1/2}} =  C \eta(z)^{11/2}
\label{eq:final_ratiowitheta}
\end{equation}

\section{Cluster Sample}
\label{sec:sample}

For our analysis, we use $Y_{SZ}$ and $Y_X$  for 58 SPT-SZ selected galaxy clusters from~\citet{bulbul19}. The SZ data have been taken from the South Pole Telescope (hereafter SPT)  which is a 10~m millimeter wave telescope located at the South Pole~\cite{Carlstrom}. One of the  main goals of the SPT is to find  galaxy clusters using their SZ signatures upto very high redshifts. SPT has imaged the sky at three different frequencies viz. 95 GHz, 150 GHz, and 220 GHz~\cite{Carlstrom}, and   carried out a 2500 square degree  survey between 2007 to 2011. A total of  516 galaxy clusters have been detected in this survey, with a mass threshold of $3\times 10^{14} M_\odot$  upto a redshift of  1.8~\cite{Bleem15,bocquet19}. Many dedicated optical surveys, e.g. DES~\cite{Saro}, BCS~\cite{Desai12} as well as targeted observations~\cite{Song} have followed-up the SPT cluster candidates in order to determine confirm these as true clusters and determine their  redshifts. 
Here, we have also used the X-ray  observations (by XMM-Newton) for  58 of these  clusters, in the redshift range 0.2 $\leq$ $z$ $\leq$ 1.5~\cite{bulbul19}. The $Y_X$ parameter from XMM-Newton measurements has been measured at $R_{500}$.
The $Y_{SZ}$ measurements provided by SPT have been obtained by averaging over  a cylindrical volume. These need to be transformed to a spherical volume with radius equal to $R_{500}$, since $Y_X$ measurements are provided at $R_{500}$. The details of the conversion procedure are discussed  in the Appendix. 

\section{Calculation of luminosity distance $D_L$}
\label{sec:calc of $D_L$}
We use three different methods to estimate the the luminosity distance, $D_L$ for each cluster. We now describe each of these methods.

\subsection{Method 1: Using a redshift cut}
We use Type Ia SNe sample from the Union 2.1 compilation~\cite{suzuki12}, which  consists of 580 Type Ia supernovae data spanning the redshift range $0.015 \leqslant z \leqslant 1.414$ (with median redshift of  $z \approx$ 0.294) in order to estimate $D_L$ for the SPT-SZ sample by choosing the closest SNe corresponding to every cluster. Similar to H19, we use the following cut on the supernova redshift ($z_{SNe}$)  for  a  cluster with redshift $z$:  $|z-z_{SNe}| \leq 0.005$. For most galaxy clusters, there are  multiple supernovae satisfying the above cut. To determine the $D_L$ and its error at the cluster location in such cases, we  first need to evaluate  the weighted average of the distance modulus of Type Ia SNe ($\bar{\mu}(z)$) and its associated error. 

To reconstruct $D_L$ from  the distance modulus $\mu$, we start from the following equation~\cite{Huterer17}:
\begin{equation}
    \mu = m-M = 5 \log_{10} \left(\frac{d_L}{10  (pc)}\right)
    \label{eq:distmod}
\end{equation}
The Type Ia SNe surveys  cannot directly measure $\mu$, as they only measure the  relative distances. The distance modulus for the Union 2.1 catalog \footnote{This catalog has been downloaded from \url{http://supernova.lbl.gov/Union/figures/SCPUnion2.1_mu_vs_z.txt}},  has obtained   by assuming $M=-19.31$ and $h=0.7$, where $h\equiv H_0/100$. Therefore, we can re-write $\mu$ as a function of $h$ as  follows: 

\begin{equation}
\mu = m-M = m - \left[-19.31 + 5 \log_{10}  \left(\frac{h}{0.7}\right)\right]
\label{eq:mumarg}
\end{equation}

Currently, there are  discrepancies between the different measurements of $H_0$, with a  fierce ongoing debate on whether this is a harbinger of new Physics or only systematics~\cite{Divalentino21}. The tension between the  model-independent low redshift based probes  and those estimated  using early time probes with the $\Lambda$CDM model ranges between 4-6$\sigma$. (See ~\cite{Divalentino21,Verde,Bethapudi}  for extensive reviews of these different measurements). \rthis{Since the $\mu$ obtained from Type Ia SNe measurements has an explicit dependence on $h$, we consider it as a nuisance parameter and marginalize over
$h$, while testing for deviations from  CDDR.
This is similar to the analysis done in ~\cite{Wu15}.
 For this purpose, we use a Gaussian prior on $h$, using  the latest $H_0$ measurement from the SH0ES team of $H_0= 73.2 \pm 1.3$ km/sec/Mpc~\cite{Riess21}.} The reason is that the SH0ES measurement is agnostic to details of the underlying $\Lambda$CDM model,  and is  based on directly measuring the distance-redshift relation by constructing a distance ladder.

One then invert Eq.~\ref{eq:distmod}
to obtain $D_L$ as a function of $h$ using the $\mu$ provided for the Union 2.1 catalog
\begin{equation}
    D_L(z) \text{(Mpc)} = \left(\frac{0.7}{h} \right)10^{(\mu - 25)/5}
    \label{eq:Dlh}
\end{equation}
In case of multiple matches for a given redshift, we use the weighted average $\bar{\mu}(z)$ in the above equation. \rthis{We then plug Eq.~\ref{eq:Dlh} in Eq.~\ref{eq:final_ratiowitheta}.}

For six clusters, we could not find any supernovae counterparts with the aforementioned redshift cut, and these clusters were  therefore culled from our sample. Thereafter, we are left with a  sample of 52 clusters for our analysis.

\subsection{Method 2: Non-parametric reconstruction  using Type Ia SNe}

\begin{figure*}
    \centering
    \includegraphics[width=10cm, height=7cm]{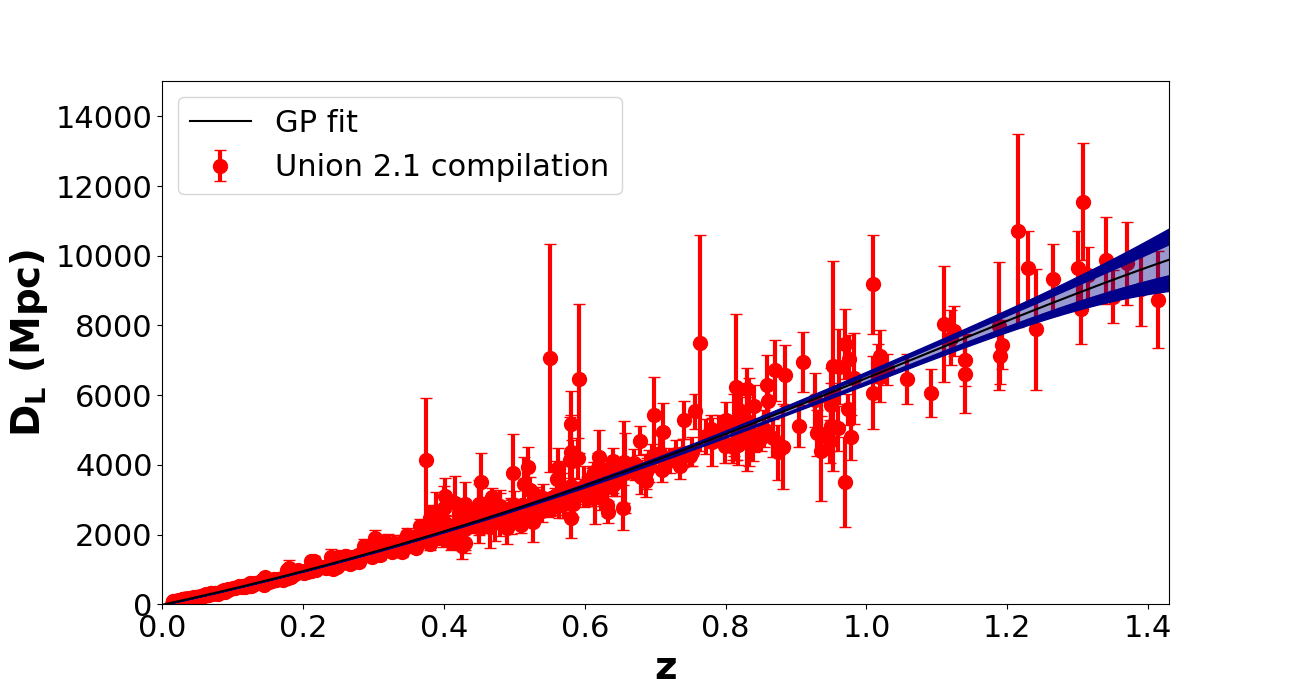}
    \caption{Reconstruction of  $D_L$
    using GPR. The red data points show the 580 Type Ia SNe taken from Union 2.1 compilation~\cite{suzuki12}. The black line indicates the best GP fit to the data along with $1\sigma$ and $2\sigma$ error bands shown by the two different shades of blue color. \rthis{For this plot we have assumed $h=0.7$.}}
    \label{fig:f2}
\end{figure*}

We now adopt another method to find $D_L$ for every cluster, by using the Gaussian Processes Regression (GPR). A Gaussian process is the generalization of a Gaussian distribution. It is characterized by a mean  and a covariance function (usually called the kernel function)~\cite{seikel12}. More details about GPR can be found in our previous works~\cite{HS,Borafg}, and references therein. For this work, we select the squared exponential (RBF) covariance function, which is given by:
\begin{equation}
 \label{eq:kernal}
  K(x,\Tilde{x}) = \sigma_f^2 \exp  {\left[\frac{-(x-\Tilde{x})^2}{2l^2}\right]}, 
\end{equation}
It depends on  two hyperparameters: $\sigma_f$ and $l$. The length parameter $l$ controls the smoothness of the kernel function. To reconstruct $D_L$ at the cluster redshift, we used the {\tt scikit-learn} module in python~\cite{sklearn}. Fig~\ref{fig:f2} shows the reconstructed luminosity distance as a function of $z$ using GPR. Using this non-parametric reconstruction, one can estimate $D_L$ for every SPT-SZ cluster. In order to be consistent with Method 1, we omit the same six clusters, for which $D_L$ could not be estimated using Method 1. \rthis{Similar to Method 1 , $D_L$ reconstructed using GPR from the  Union 2.1 sample  also needs  to be rescaled as a function of $h$,  while testing for CDDR violation, in order to marginalize over $h$.}

\subsection{Method 3: Reconstruction  using cosmic chronometers}

\begin{figure*}
    \centering
    \includegraphics[width=10cm, height=7cm]{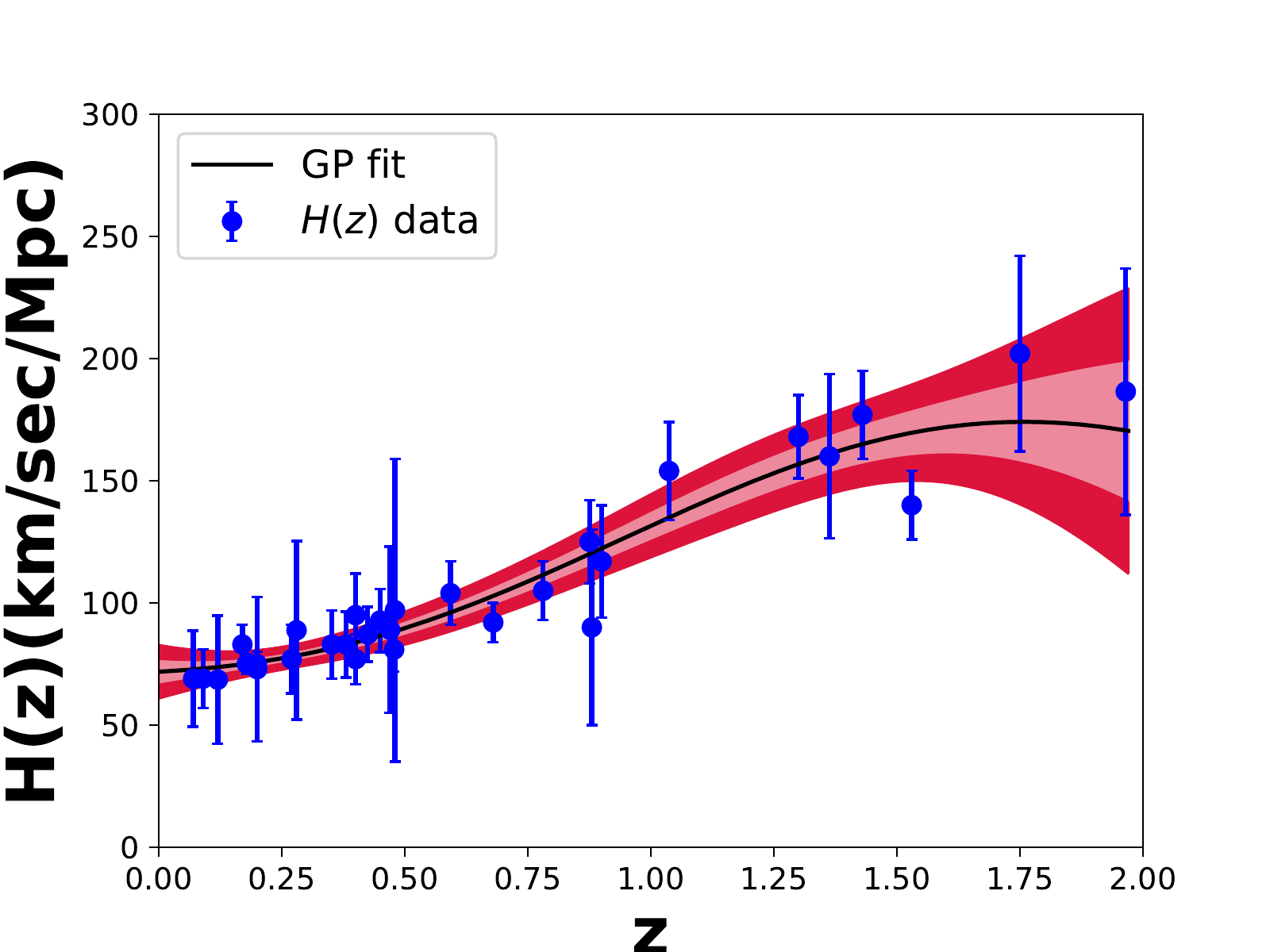}
    \caption{Reconstruction of $H(z)$ using GPR. The blue data points represent the 31 $H(z)$ cosmic chronometer measurements  compiled by~\citet{li19}. The black line indicates the best GP fit to data along with $1\sigma$ and $2\sigma$ error bands shown by the two different shades of red color.}
    \label{fig:f3}
\end{figure*}

Here, we used 31 $H(z)$ measurements from~\citet{li19} (same as that used in~\cite{HS,Borafg}) spanning the redshift range $0.07 \leqslant z \leqslant 1.965$.  These $H(z)$ measurements are obtained from cosmic chronometers, which provide a model agnostic estimate of the expansion history at any redshift $z$~\cite{Jimenez}.
Using these $H(z)$ measurements,   we have again used the {\tt scikit-learn} module for GPR, to estimate $H(z)$ for any redshift. Furthermore, the reconstructed $H(z)$ data can be used to derive the luminosity distance $D_L$ which can be written as,
\begin{equation}
D_{L} (z) = c (1+z)\int_{0}^{z}\frac{dz^{'}}{H(z')}
\label{eq:dl}
\end{equation}
where $H(z')$ is the reconstructed Hubble data. Fig~\ref{fig:f3} shows the reconstructed $H(z)$ data from chronometers as a function of $z$ along with their  $1\sigma$ and $2\sigma$ uncertainties.   Note that, here too, the same six clusters as in Method 1 have been omitted.

\begin{figure*}
    \centering
    \includegraphics[width=10cm, height=7cm]{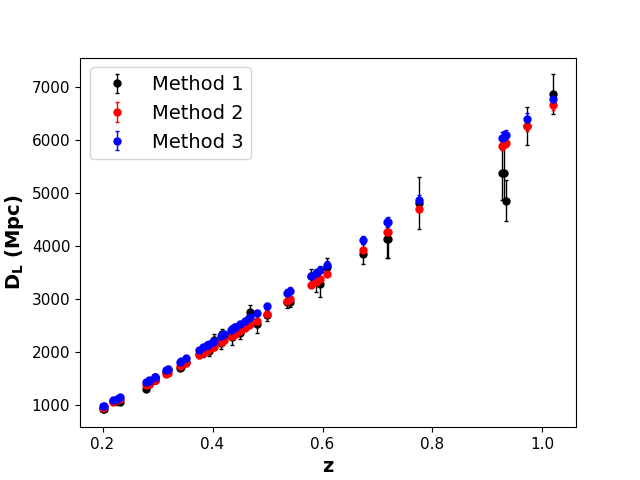}
    \caption{A comparison of  $D_L$ for the  SPT cluster sample  obtained from the different methods as discussed in Sec~\ref{sec:calc of $D_L$}.  The fractional median difference between the luminosity distance $D_L$ obtained by Method 1, 2, and 3, when compared against each other is equal to  2.3\%, 5.4\%, 2.9\% respectively and hence is negligible. \rthis{Note that the  $D_L$ values  for  Method 1 and 2 in this plot  are obtained by assuming $h=0.7$.}}
    
    \label{fig:f4}
\end{figure*}

A comparison plot of the luminosity distance $D_L$ as a function of $z$ using  all the three methods for the 52 SPT-SZ clusters is shown in Fig~\ref{fig:f4}. The fractional median difference between the luminosity distance $D_L$ obtained by Method 1, 2, and 3, when compared against each other is equal to 2.3\%, 5.4\%, 2.9\% between Method 1-2,  Method 2-3, and  Method 1-3  respectively. Therefore, this difference in $D_L$ between the three methods is marginal.

\section{Analysis and Results}
\label{sec: analysis and results}
In order to constrain $\eta(z)$ (as used in Eq.~\ref{eq:final_ratiowitheta}), we use the following parametric forms~\cite{hees14}. These were also used in H19 and other previous works testing for a violation of CDDR.

\begin{eqnarray}
     \eta(z) &=& 1+\eta_0 z
     \label{eq:P1} \\
     \eta(z) &=& 1+\eta_0 [z/(1+z)]
     \label{eq:P2} \\
    \eta(z) &=& 1 + \eta_0 \ln(1+z)
     \label{eq:P3} \\
    \eta(z) &=& (1+z)^{\eta_0}
     \label{eq:P4}
\end{eqnarray}

\begin{figure*}
    \centering
    \includegraphics[width=20cm, height=14cm]{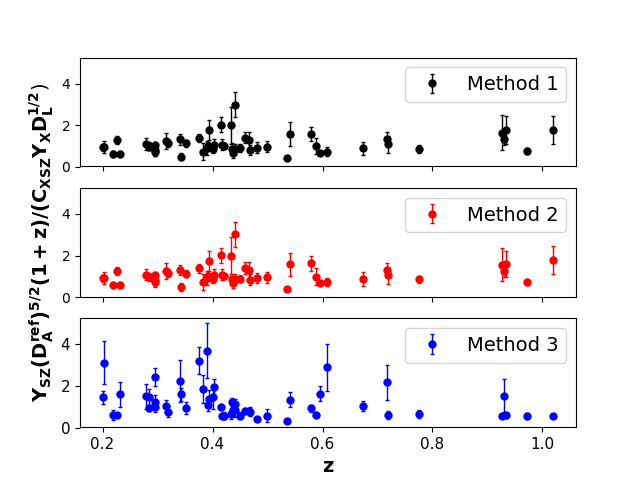}
     \caption{A comparison of the $Y_{SZ}$-$Y_X$ ratio defined in Eq.~\ref{eq:final_ratiowitheta}  for the three  different methods used for calculating $D_L$, as  described in Sec.~\ref{sec:calc of $D_L$}. \rthis{Note that for making this plot for  Method 1 and Method 2, we have used  $h=0.7$.} }
    \label{fig:f5}
\end{figure*}

For all the above parametrizations, $\eta_0$ encapsulates a possible departure from the standard CDDR, and  $\eta_0 = 0$ corresponds to no CDDR violation. Table~\ref{tab:possible combination} summarizes  all possible combinations  of the different methods used for $D_L$ measurements and  the multiple $\eta(z)$ parametrizations,  we have analyzed in this work. The $Y_{SZ}-Y_X$ ratio given by Eq.\ref{eq:final_ratiowitheta} is shown  in Fig~\ref{fig:f5}  for each of the methods used for reconstructing $D_L$ (as discussed in Sec.~\ref{sec:calc of $D_L$}).

We now proceed to obtaining constraints on $\eta(z)$ for each of the parametrizations. The likelihood function $\mathcal{L}$ used in our analysis is given by:
\begin{equation}
  -2 \ln \mathcal{L}  =  \sum_{i=1}^{N}\frac{(\eta_{obs}(z_i)- C\eta(z)^{11/2})^2}{\sigma_{i}^2} + \sum_{i=1}^{N} \ln 2\pi{\sigma_{i}^2} ,
 \label{eq:likelihood} 
\end{equation} 
where $\eta_{obs}$ is obtained from Eq.~\ref{eq:final_ratiowitheta} and $\sigma_i$ denotes the total error calculated as follows:
\begin{equation}
     \sigma_{i}^2 =  \sigma_{\eta}^2  +  \sigma_{int}^2
     \label{eq:sigma}
\end{equation}
where, $\sigma_{\eta}^2$ represents the error  in $\eta_{obs}$ which is calculated by propagating the errors in $Y_{SZ}$,$Y_X$, $D_L(z)$, and $D_A^{ref}$. \rthis{When we estimate $D_L$ using  Method 1 and Method 2, it will also have an additional dependence on $h$ (cf. Eq.~\ref{eq:Dlh}). Therefore, any estimate on the violation of CDDR using the $D_L$ obtained from these two methods will contain an additional free parameter, i.e., $h$ which has to be marginalized over in order to constrain the CDDR parameters. As mentioned earlier, we use a Gaussian prior on $h$, with a   mean equal to 0.732 and scale parameter of 0.0013}
We have also included an intrinsic scatter term ($\sigma_{int}$)  as a free parameter, which is added in quadrature to the observational error,  while maximizing the likelihood.  We used the  {\tt emcee} MCMC sampler~\cite{emcee} to estimate the model parameters i.e. $\eta_0$ and $C$, by maximizing the likelihood function defined in Eq.~\ref{eq:likelihood}.

\begin{figure*}
    \centering
    \includegraphics[width=10cm,height=8cm]{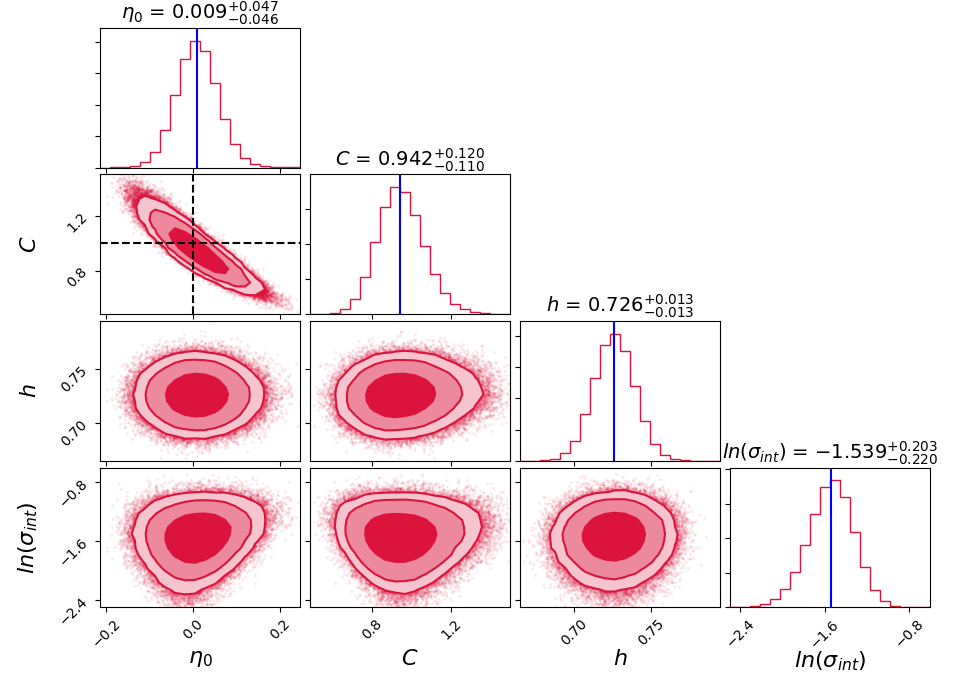}
   \caption{\textbf{Case I:} Constraints on the parameters $\eta_0$, $C$, \rthis{and $h$} along with $\ln(\sigma_{int})$. The plots along the diagonal are the one-dimensional marginalized likelihood distributions. The contour plot represents the two-dimensional marginalized constraints showing the 68\%, 95\%, and 99\% credible regions. These contours have been obtained using the {\tt Corner} python module~\cite{corner}.}
    \label{fig:f6}
\end{figure*}

\begin{figure*}
    \centering
    \includegraphics[width=10cm,height=8cm]{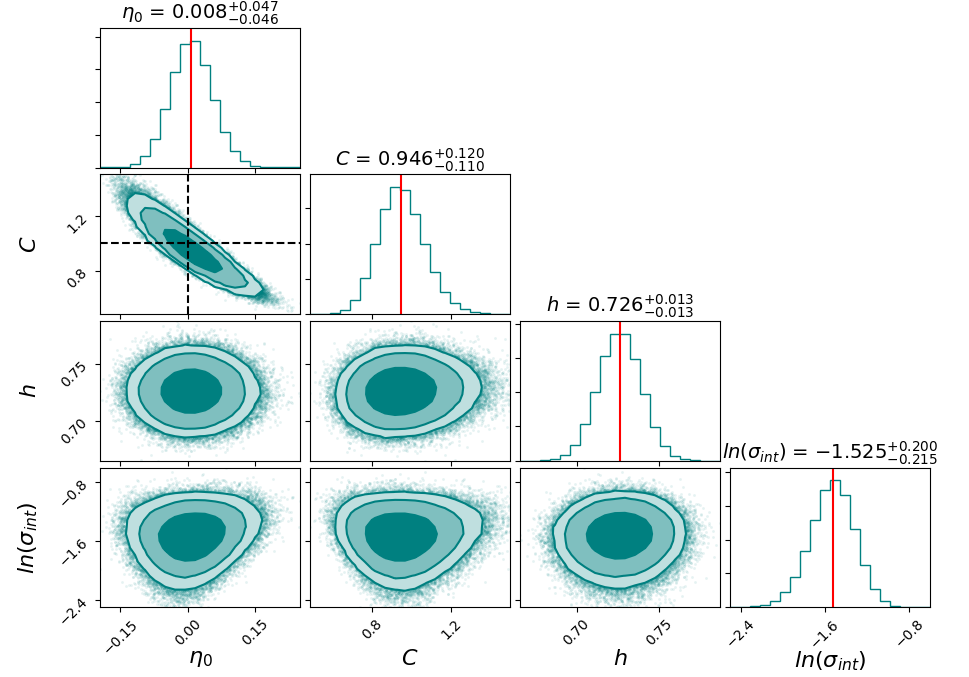}
    \caption{\textbf{Case V:} Constraints on the parameters $\eta_0$, $C$, \rthis{and $h$} along with $\ln(\sigma_{int})$. The plots along the diagonal are the one-dimensional marginalized likelihood distributions. The contour plot represents the two-dimensional marginalized constraints showing the 68\%, 95\%, and 99\% credible regions. These contours have been obtained using the {\tt Corner} python module~\cite{corner}. }
    \label{fig:f7}
\end{figure*}

\begin{figure*}
    \centering
    \includegraphics[width=10cm,height=8cm]{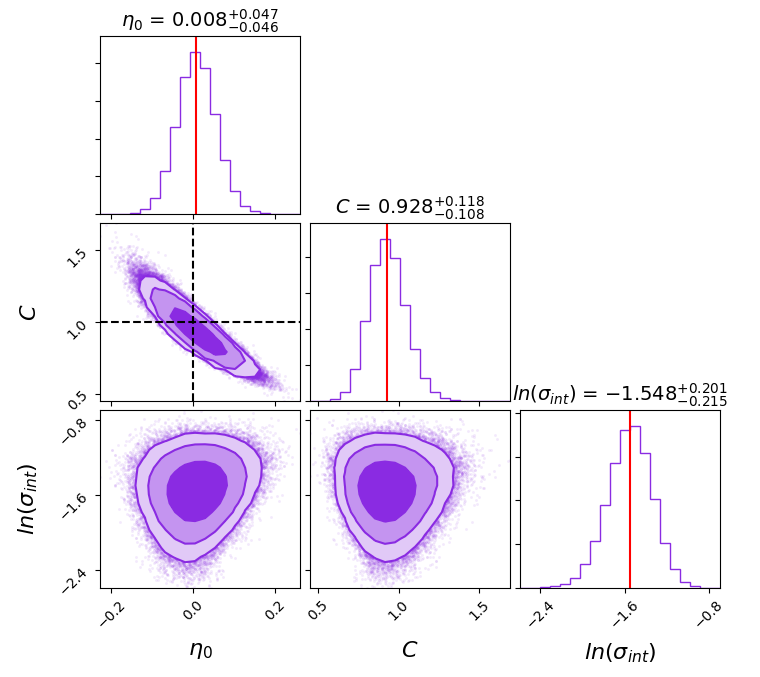}
    \caption{\textbf{Case IX:} \rthis{ Constraints on the parameters $\eta_0$, and $C$ along with $\ln(\sigma_{int})$. The plots along the diagonal are the one-dimensional marginalized likelihood distributions. The contour plot represents the two-dimensional marginalized constraints showing the 68\%, 95\%, and 99\% credible regions. These contours have been obtained using the {\tt Corner} python module~\cite{corner}. Note that unlike Fig.~\ref{fig:f6} and Fig.~\ref{fig:f7}, there is no contour for $h$, since $D_L$ estimated using this method (Method 3) does not  depend on $h$.}}
    \label{fig:f8}
\end{figure*}

\begin{figure*}
    \centering
    \includegraphics[width=15cm,height=9cm]{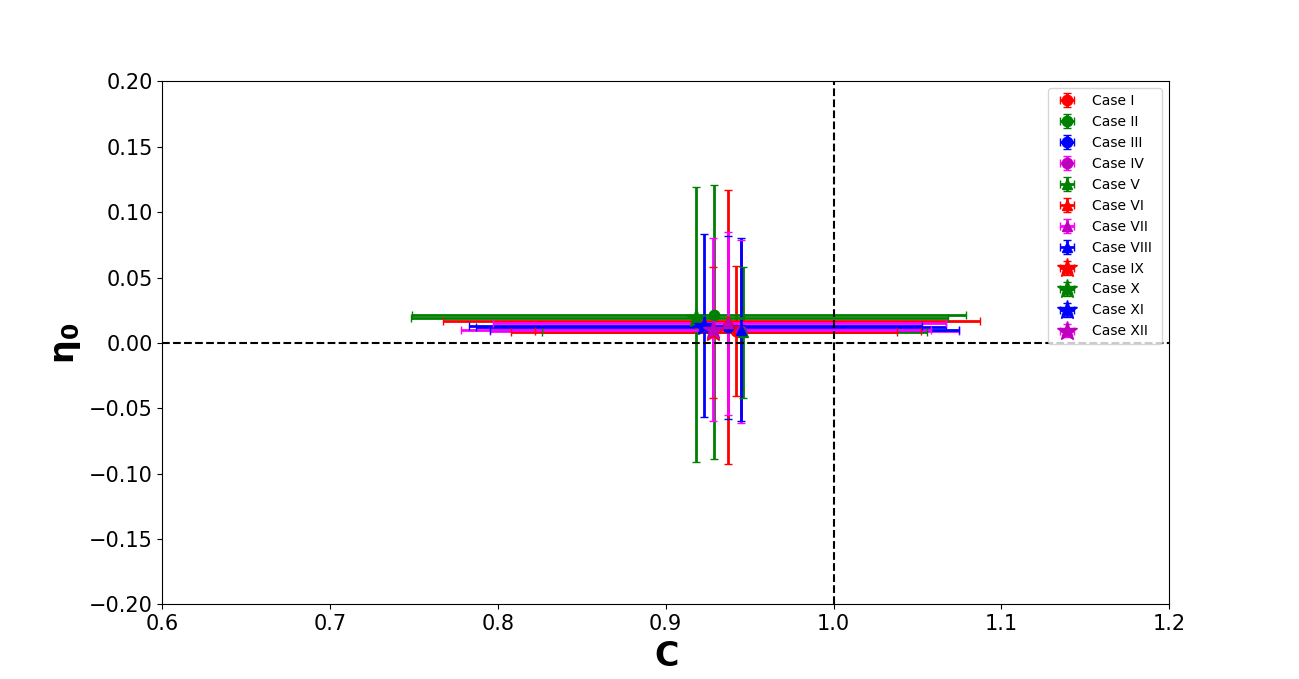}
    \caption{A comparison of different cases studied in this work. The black dashed lines $(C,\eta_0)$ = (1,0) correspond to the standard distance duality relation. A combination of different colors with different data style represent all different cases used in this study. For all the analyses carried out,  $\eta_0$ is consistent with zero. }
    \label{fig:f10}
\end{figure*}

The marginalized one-dimensional posteriors for each of the parameters along with the 2-D 68\%, 95\%, and 99\%  credible intervals, are displayed in Fig.~\ref{fig:f6}-~\ref{fig:f8} for   Eq.~\ref{eq:P1}, \rthis{corresponding  to Methods 1, 2, and 3, respectively for calculating $D_L$. Note that in Fig~\ref{fig:f6} and Fig.~\ref{fig:f7},  we need to marginalize over $h$, since $D_L$ has been estimated using Method 1 and Method 2. However, there is no dependence on $h$ in Method 3,  and hence Fig.~\ref{fig:f8} does not  contain any contours for  $h$.}
A complete  summary of all our results for each of the four parametric forms using all the  three methods for computing $D_L$  is tabulated in Table~\ref{tab:comparison_table}. This table also contains a summary of previous results in literature for the same parametric forms, whenever available. A graphical summary of all our results for the different cases can be found  in Fig~\ref{fig:f10}.
Therefore, the results from all these analyses indicate that there is no violation of CDDR, and the parameter which encapsulates any violation of CDDR as a function of redshift is consistent with zero to within 1$\sigma$ for all the parametric forms used.


\begin{table*}

	\centering
	\begin{tabular}{|l|c|c|c|c|c|c|c|r|} 
		
 \hline
		
		\textbf{Combination} &  	\textbf{Method 1} & \textbf{Method 2} & \textbf{Method 3} & \textbf{Eq.~\ref{eq:P1}} & \textbf{Eq.~\ref{eq:P2}} & \textbf{Eq.~\ref{eq:P3}} & \textbf{Eq.~\ref{eq:P4}}  \\ 
           
		\hline
		Case I & \checkmark & & &  \checkmark & & & \\
	    Case II & \checkmark & & & & \checkmark & &\\
		Case III  & \checkmark & & & & &\checkmark &  \\
	    Case IV & \checkmark & & & & & & \checkmark \\
		Case V & &\checkmark & &  \checkmark & & & \\
	    Case VI & &\checkmark & & &  \checkmark & &\\
		Case VII  & &\checkmark & &  & &\checkmark &  \\
	    Case VIII & &\checkmark & &  & & & \checkmark \\
	    Case IX & & &\checkmark  &  \checkmark & & & \\
	    Case X & & &\checkmark  & &  \checkmark & &\\
		Case XI  & & &\checkmark &   & &\checkmark &  \\
	    Case XII & & &\checkmark &  & & & \checkmark \\
		\hline
	\end{tabular}
\caption{An illustration of the different analyses  used for testing CDDR.}
\label{tab:possible combination}
\end{table*}

\begin{table*}
	\centering
	\begin{tabular}{|l|c|c|c|c|c|r|} 
	\hline
	\textbf{Dataset used}  &  	\textbf{Eq.~\ref{eq:P1}} & \textbf{Eq.~\ref{eq:P2}} & \textbf{Eq.~\ref{eq:P3}} & \textbf{Eq.~\ref{eq:P4}} & \textbf{Reference} \\
	\hline
	Angular Dia Dist+SNe Ia & $-0.28\pm0.44$($2\sigma$) & $-0.43\pm0.60$($2\sigma$) & - & - & ~\cite{holanda10}\\
	Angular Dia Dist+SNe Ia & $-0.15\pm0.17$ & $-0.23\pm0.24$ & - & - & ~\cite{xiang11}\\
	Angular Dia Dist+SNe Ia & $-0.07\pm0.19$ & $-0.11\pm0.26$ & - & - & ~\cite{li11}\\
   $f_{gas}$ & $-0.06\pm0.16$ & $-0.07\pm0.24$ & - & - & ~\cite{holanda12}\\
    $f_{gas}$+SNe Ia & $-0.03_{-0.65}^{+1.03}$ & $-0.08_{-1.22}^{+2.28}$ & - & - & ~\cite{goncalves12}\\
    Angular Dia Dist+SNe Ia & $-0.23\pm0.23$($2\sigma$) & $-0.35\pm0.37$($2\sigma$) & - & - & ~\cite{liang13}\\
	Angular Dia Dist+SNe Ia & $0.16_{-0.39}^{+0.56}$ & - & - & - & ~\cite{yang13}\\
	Angular Dia Dist+$H(z)$ & $-0.10_{-0.13}^{+0.12}$ & $-0.16_{-0.19}^{+0.18}$ & - & - & ~\cite{costa15} \\
	$f_{gas}$+$H(z)$ & $0.062_{-0.15}^{+0.17}$ & $-0.17_{-0.28}^{+0.34}$ & - & - & ~\cite{costa15}\\
	Angular Dia Dist+$f_{gas}$+SNe Ia+$T_{CMB}$ & $-0.012\pm0.022$ & $-0.02\pm0.034$ &$-0.017\pm0.027$  & $-0.017\pm0.026$ & ~\cite{holanda1705} \\
	Angular Dia Dist+$f_{gas}$+SNe Ia+$T_{CMB}$ & $-0.011\pm0.021$ & $-0.015\pm0.033$ & $-0.013\pm0.027$  & $-0.013\pm0.028$ &~\cite{holanda1705} \\
	Strong Grav. Lensing+SNe Ia+GRBs & $0.00\pm0.10$($2\sigma$) & $-0.36_{-0.42}^{+0.37}$($2\sigma$)  & $-0.10\pm0.24$($2\sigma$) & $-0.16_{-0.51}^{+0.24}$($2\sigma$)  & ~\cite{holanda1611} \\
	Strong Grav. Lensing+SNe Ia+GRBs & $0.15\pm0.13$($2\sigma$) & $-0.18_{-0.65}^{+0.45}$($2\sigma$)  & $0.22_{-0.32}^{+0.40}$($2\sigma$)  & $0.27_{-0.38}^{+0.22}$($2\sigma$)  & ~\cite{holanda1611} \\
	
	$Y_{SZ}-Y_X$+SNe Ia & $0.05\pm0.07$($2\sigma$) & $0.09\pm0.16$($2\sigma$) & - & - & ~\cite{holanda19}\\
	$\mathbf{Y_{SZ}-Y_X}$ \textbf{ratio+SNe Ia(Method 1)} & \rthis{$\mathbf{0.009\pm0.05}$} & \rthis{$\mathbf{0.021\pm0.11}$} & \rthis{$\mathbf{0.012\pm0.07}$} & \rthis{$\mathbf{0.009\pm0.07}$} & \textbf{This work}\\
	
	$\mathbf{Y_{SZ}-Y_X}$ \textbf{ratio+SNe Ia(Method 2)} & \rthis{$\mathbf{0.008\pm0.05}$} & \rthis{$\mathbf{0.017\pm0.11}$} & \rthis{$\mathbf{0.015\pm0.07}$} & \rthis{$\mathbf{0.010\pm0.07}$} & \textbf{This work}\\
	
	$\mathbf{Y_{SZ}-Y_X}$ \textbf{ratio+}$\boldsymbol{H(z)}$ \textbf{(Method 3)} & \rthis{$\mathbf{0.008\pm0.05}$} & \rthis{$\mathbf{0.019\pm0.11}$} & \rthis{$\mathbf{0.013\pm0.07}$} & \rthis{$\mathbf{0.010\pm0.07}$} & \textbf{This work}\\
    \hline
    
	\end{tabular}
\caption{Constraints on $\eta_0$ from previous studies using different parametric forms as defined in Eq.~\ref{eq:P1} - Eq.~\ref{eq:P4} along with our results presented in last three rows. The quoted uncertainties are at $1\sigma$ wherever not mentioned explicitly. We do not find any violation of CDDR from our analyses.}
\label{tab:comparison_table} 
\end{table*}

\section{Conclusions}
\label{sec:conclusion}
In this work, we carried out a model-independent test  of CDDR, along the same lines as H19. For this purpose,  we used  52 SPT-SZ clusters  in the redshift range 0.2 $\leq$ $z$ $\leq$ 1.5 in conjunction  with X-ray measurements from XMM-Newton~\cite{bulbul19}. To get independent distance measurements at the cluster redshifts, we used  $D_L$ from Union 2.1 Type Ia SNe sample ~\cite{suzuki12} and $H(z)$ measurements from cosmic chronometers~\cite{li19}. The cluster observables used to test CDDR  include the SZ Compto-ionization parameter ($Y_{SZ}$),  along with its X-ray counterpart ($Y_X$), both of which are independent thermal energy proxies, having  different scalings with $\eta(z)$ (cf. Eq.~\ref{eq:Y_SZDa2} and ~\ref{eq:Y_x}). The $Y_{SZ}$ to $Y_X$ ratio is then used to test the CDDR, which is given by Eq.~\ref{eq:final_ratiowitheta}. One input needed for this analysis is the luminosity distance at the cluster redshift, for which we  have used three  independent estimators. These are  summarized in Sect~\ref{sec:calc of $D_L$} and tabulated in Table~\ref{tab:possible combination}. 

The dimensionless ratio of $Y_{SZ}$ to $Y_X$ (after suitable scalings)  is shown in Fig~\ref{fig:f5} as a function of $z$ for all the three methods used for estimating $D_L$.

To test the validity of CDDR, we chose different parametric forms, defined in Eqs.~\ref{eq:P1} - Eq.~\ref{eq:P4}, where the parameter $\eta_0$ characterizes any departure from CDDR.
Table~\ref{tab:possible combination} shows all the possible used cases studied in this work.    
Furthermore, to estimate the parameters in Eqs.~\ref{eq:P1} - Eq.~\ref{eq:P4}, we maximize the likelihood function given by Eq~\ref{eq:likelihood}.  \rthis{Two of the three methods obtain the $D_L$ value  from the Union 2.1 catalog, which assumed $h=0.7$. While using these $D_L$ values, we  marginalized over $h$, using a Gaussian  prior  over $h$ from the latest SH0ES measurement~\cite{Riess21}, in order to obtain marginalized constraints on the CDDR violating parameters.}

Our results for all these analyses  can be found in  Table~\ref{tab:comparison_table} along with a comparison with  previous results.
A graphical summary of all our results for CDDR (for all the cases studied herein) can be found in Fig~\ref{fig:f10}. \rthis{For Cases I, V, and IX, which test the parametric form in  Eq.~\ref{eq:P1} using Methods 1, 2, and 3, respectively} the joint 2-D credible intervals for $\eta_0$,  $C$ (a parameter in the $Y_{SZ}$ to $Y_{X}$ ratio) and $h$ (if needed)  can be found  in Fig~\ref{fig:f6} - Fig~\ref{fig:f8}. Our results from all these analyses  are consistent with  CDDR within $1\sigma$.
Therefore, we conclude that  there is no  evidence for a violation of CDDR.

\section*{APPENDIX}

The $Y_{SZ}$ measurements provided by SPT have been obtained by averaging over  a cylindrical   volume  within an aperture radius of $0.75R'$~\cite{Bleem15}. Since, the $Y_{SZ}$ parameter defined in Eq.~\ref{eq:Y_SZDa2} needs to be averaged over a  spherical  volume, we need to convert the cylindrically averaged $Y_{SZ}$ (provided by SPT)  to a  spherically averaged $Y_{SZ}$ at $R_{500}$, for which we have used the following conversion method~\cite{arnaud10,melin11}:

\begin{widetext}
\begin{eqnarray}
Y_{cyl}(R_1) &=& Y_{sph}(R_b) - \frac{\sigma_T}{m_e c^2} \int_{R_1}^{R_b} 4\pi P(r) \sqrt{r^2-R_1^2}r dr 
\label{eq:cylY} \\
Y_{sph}(R_2) &=& \frac{\sigma_T}{m_e c^2} \int_{0}^{R_2} 4\pi P(r) r^2 dr
\label{eq:sphY}
\end{eqnarray}
\end{widetext}
where the subscripts $cyl$ and $sph$ denote the $Y_{SZ}$ parameter,  measured in cylindrical and spherical  volumes, respectively. $R_1$ indicates the cylindrical aperture in  which $Y_{cyl}$ is measured and  $R_b$ denotes the radial extent of the cluster. We assume $R_b$ = $10R_{500}$~\cite{arnaud10}, $R_1 = 0.75'R_{500}$ corresponding to the SPT measurement, and $R_2 = R_{500}$, since $Y_X$ measurements are provided at $R_{500}$. Note that in ~\cite{Liu}, the SPT collaboration assumed $R_b=5 R_{500}$, in order to convert $Y_{SZ} (cyl)$ to $Y_{SZ} (sph)$, for a sample of low-mass clusters and groups.
However, the difference in the estimation of $Y_{SZ} (sph)$ with these two values of $R_b$ is negligible.
In Eq.~\ref{eq:sphY}, $P(r)$ refers to the pressure profile.
To do the conversion, we have used the Universal Pressure Profile (UPP) for $P(r)$~\cite{arnaud10}, similar to our previous work~\cite{bora21}.
\begin{equation}
    \frac{P(r)}{P_{500}} = \frac{P_0}{x^{\gamma}(1+x^{\alpha})^{(\beta-\gamma)/\alpha}} 
\label{eq:UPP}
\end{equation}
where $x=r/R_{500}$, and $P_{500}$ is given by~\cite{Nagai07}
\begin{equation}
    P_{500} = 1.65\times10^{-3} E(z)^{8/3} \left[\frac{M_{500}}{3\times10^{14} h_{70}^{-1} M_\odot}\right]^{2/3} h^2_{70}   keV cm^{-3}
\end{equation}

\begin{figure*}
    \centering
    \includegraphics[scale =0.45]{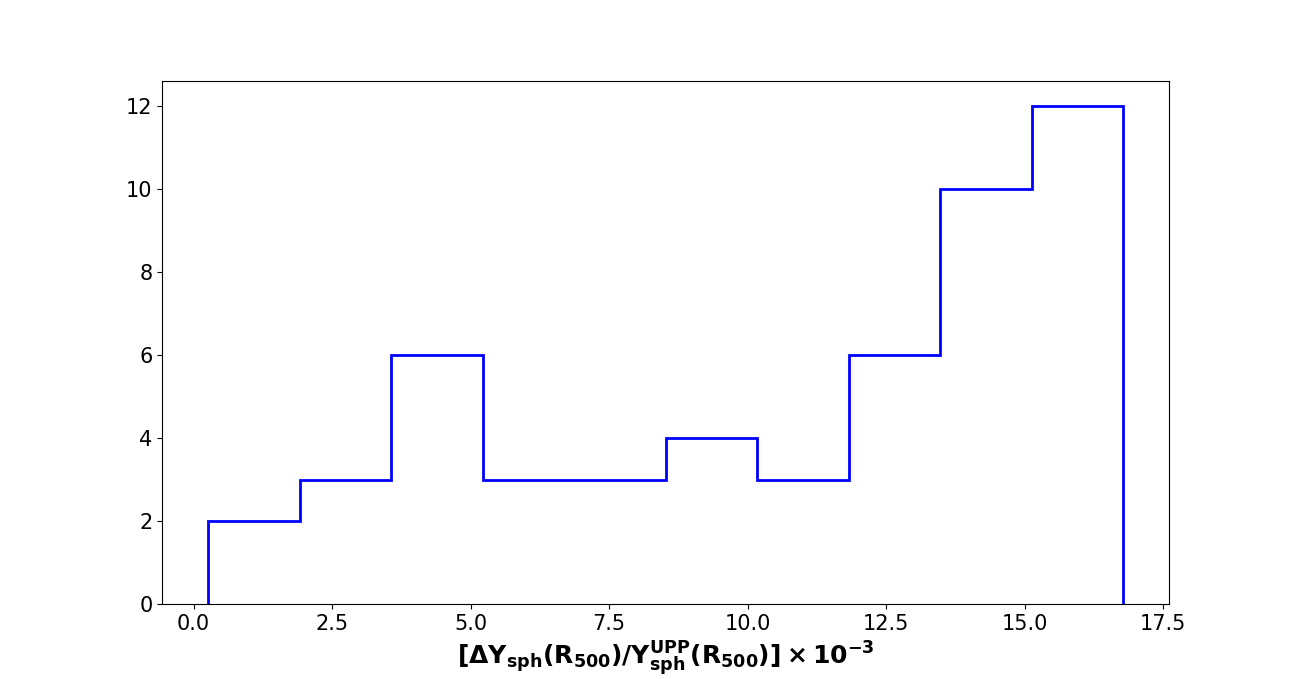}
    \caption{Histogram of the fractional deviation in $Y_{sph}(R_{500})$ between the  UPP and BPP pressure profile for  the SPT-SZ clusters analyzed in this work. As we can see, the fractional  deviation is less than 1\%.} 
    \label{fig:hist}

\end{figure*}

As a cross-check we also did the conversion using the Battaglia et al (BPP) pressure profile~\cite{battaglia10}, which is given by
\begin{equation}
     P/P_{500} = A[1+(x/x_c)^{\alpha}]^{-\gamma/\alpha}
     \label{eq:BPP}
\end{equation}
where $x=r/R_{500}$; and the values of the constants $A$, $\alpha$, and $\gamma$ are provided in ~\cite{battaglia10}. A histogram of the  difference in $Y_{SZ} (sph)$ between the two pressure profiles is shown in Fig.~\ref{fig:hist}. Since the difference between the two estimates is negligible, we  do only report  results using the UPP profile in this manuscript.

Therefore, from Eq.~\ref{eq:cylY} and Eq.~\ref{eq:sphY},  we can estimate the ratio of $Y_{sph}(R_{500})$ to $Y_{cyl}(0.75' D_A)$. By using this ratio one can  estimate  $Y_{sph}(R_{500})$ for every SPT-SZ cluster.

\section*{ACKNOWLEDGEMENT}

KB would like to acknowledge the Department of Science and Technology, Government of India for providing the financial support under DST-INSPIRE Fellowship program. We are also grateful Nao Suzuki for important clarifications about the Union 2.1 sample and  the anonymous referee for useful constructive feedback on our manuscript.

\bibliography{ref} 

\end{document}